\documentclass[final,english]{bullsrsl}[2022/06/15]

\usepackage[latin1]{inputenc}
\usepackage[T1]{fontenc}
\usepackage{natbib} 
\usepackage{graphicx}

\begin{document}
\title{Serendipitous Detection of Orbital Debris by the International Liquid Mirror Telescope: First Results}

\author[affil={1}, corresponding]{Paul}{Hickson}
\author[affil={2,3}]{Bhavya}{Ailawadhi}
\author[affil={4,5}]{Talat}{Akhunov}
\author[affil={6}]{Ermanno}{Borra}
\author[affil={2,7}]{Monalisa}{Dubey}
\author[affil={2,7}]{Naveen}{Dukiya}
\author[affil={1}]{Jiuyang}{Fu}
\author[affil={1}]{Baldeep}{Grewal}
\author[affil={2}]{Brajesh}{Kumar}
\author[affil={2}]{Kuntal}{Misra}
\author[affil={2,3}]{Vibhore}{Negi}
\author[affil={2,8}]{Kumar}{Pranshu}
\author[affil={1}]{Ethen}{Sun}
\author[affil={9,10}]{Jean}{Surdej}

\affiliation[1]{Department of Physics and Astronomy, University of British Columbia, 6224 Agricultural Road, Vancouver, BC V6T 1Z1, Canada}
\affiliation[2]{Aryabhatta Research Institute of Observational sciencES (ARIES), Manora Peak, Nainital, 263001, India}
\affiliation[3]{Department of Physics, Deen Dayal Upadhyaya Gorakhpur University, Gorakhpur, 273009, India}
\affiliation[4]{National University of Uzbekistan, Department of Astronomy and Astrophysics, 100174 Tashkent, Uzbekistan}
\affiliation[5]{ Ulugh Beg Astronomical Institute of the Uzbek Academy of Sciences, Astronomicheskaya 33, 100052 Tashkent, Uzbekistan}
\affiliation[6]{Department of Physics, Universit\'{e} Laval, 2325, rue de l'Universit\'{e}, Qu\'{e}bec, G1V 0A6, Canada}
\affiliation[7]{Department of Applied Physics, Mahatma Jyotiba Phule Rohilkhand University, Bareilly, 243006, India}
\affiliation[8]{Department of Applied Optics and Photonics, University of Calcutta, Kolkata, 700106, India}
\affiliation[9]{Institute of Astrophysics and Geophysics, University of Li\`{e}ge, All\'{e}e du 6 Ao$\hat{\rm u}$t 19c, 4000, Li\`{e}ge, Belgium}
\affiliation[10]{Astronomical Observatory Institute, Faculty of Physics, Adam Mickiewicz University, ul. Sloneczna 36, 60-286 Poznan, Poland}

\correspondance{hickson@physics.ubc.ca}
\date{29th April 2023}
\maketitle

% \author[affil1]{FirstName (+ MiddleInitials if necessary)}{FamilyName}
% \author[affil2]{...}{}
% \equalcontribauthor[]{}{} % Maximum two --> counter
% \consortium[affil]{Consortium Name}
% With consortium: affiliation will be set to "See Appendix 1 for a full
% list of consortium members and their respective affiliations
% \affiliation[affil1]{...}
% \affiliationq[affil2]{...}

% \correspondence[]{}
% No explicit corresponding author: use first author
% 

% Abstract of the paper in the same language as the paper
\begin{abstract}
Orbital debris presents a growing risk to space operations, and is becoming a significant source of contamination of astronomical images. Much of the debris population is uncatalogued, making the impact more difficult to assess. We present initial results from the first ten nights of commissioning observations with the International Liquid Mirror Telescope, in which images were examined for streaks produced by orbiting objects including satellites, rocket bodies and other forms of debris. We detected 83 streaks and performed a correlation analysis to attempt to match these with objects in the public database. 48\% of these objects were uncorrelated, indicating substantial incompleteness in the database, even for some relatively- bright objects. We were able to detect correlated objects to an estimated magnitude of 14.5 and possibly about two magnitudes greater for the faintest uncorrelated object.
\end{abstract}


\section{Introduction}

The Earth-orbit environment is becoming increasingly crowded. More than 27,000 resident space objects (RSOs) are presently catalogued, which include satellites, rocket boosters and orbital debris. It is expected that there are many more undetected objects that pose a significant risk to space operations, particularly in low-earth orbit \citep{Kessler_1978, Liou_2006}. The advent of constellations containing tens of thousands of satellites will greatly increase the numbers of resident space objects. The astronomical impact of these objects has long been recognized, but contamination of astronomical images by satellite tracks is now becoming increasingly problematic \citep{Boley_2021, Lawler_2021, Shara_1986, Hainaut_2020, Kruk_2023}.

The International Liquid Mirror Telescope (ILMT) is a 4-m zenith-pointing optical telescope located at Devasthal Peak in India \citep[$29.36^\circ$ North latitude, ][]{Surdej_2018}. Its 16-MPixel CCD gives a $0.373^\circ\times 0.373^\circ$ field of view. In order to compensate for image motion due to the Earth's rotation, the CCD is operated in time-delay integration mode in which it is continuously scanned at the sidereal rate. The telescope saw first light in April 2022 and began a period of commissioning in October.

The ILMT provides a unique opportunity to serendipitously monitor the orbital environment \citep{Pradhan_2019}. On average approximately 100 catalogued objects pass through the field of view of the ILMT each day. Typically about 6\% of these transit during dark hours while also being illuminated by the Sun and can potentially be detected by the telescope. We present here an analysis of ten nights of engineering observations obtained with the ILMT in October and November 2022.

\section{Observations and analysis}

The data set that we employed consists of 515 images obtained on the nights of October 23 to November 1, 2022, inclusive. The effective integration time for celestial objects is 102 s, which is the time that it takes for the image to drift the length of the CCD. The integration time for RSOs is usually much less as they generally move at high angular rates. These images were preprocessed and then astrometrically and photometrically calibrated using Gaia stars in the field. Background variations were removed by high-pass median filtering and bright stars (G < 18) were removed. The images were searched visually for linear tracks. The detected tracks were then measured to determine the length, width, orientation and integrated flux.

A complete set of publicly-available two-line elements (TLEs), for objects tracked by the U. S. Space Surveillance Network, was downloaded from Space-Track.org, for the period extending 30 days before and after the observations. A list of ``current'' TLEs was then generated for each of the 10 observation nights by selecting, for every cataloged object, the TLE that had the closest epoch to that of the time of observation. This formed our comparison data set. Each TLE was then propagated, using the SGP4/SDP4 algorithm \citep{Hoots_1980, Vallado_2006}, and the times and orbital parameters for all objects passing within 0.3 degrees of the zenith were extracted. A calculation was performed to determine which objects were illuminated by the sun at those times. A comparison was then made of the catalogue lists and the detected streaks. The data analysis and TLE propagation was performed using the OCS software package \citep{Hickson_2019}.

\section{Results}

Our results are summarized in Table\,\ref{tab:results}. A total of 83 streaks were identified in the ILMT images. 52\% of these were correlated with catalogued objects, based on position, time of transit and position angle. Magnitudes of the correlated objects were estimated from the measured integrated flux and the integration time found by dividing the track length by the angular rate determined from the TLE. Magnitudes cannot be determined for uncorrelated objects as the angular rates are unknown.

\begin{figure}[t]
\centering
\includegraphics[width=\textwidth]{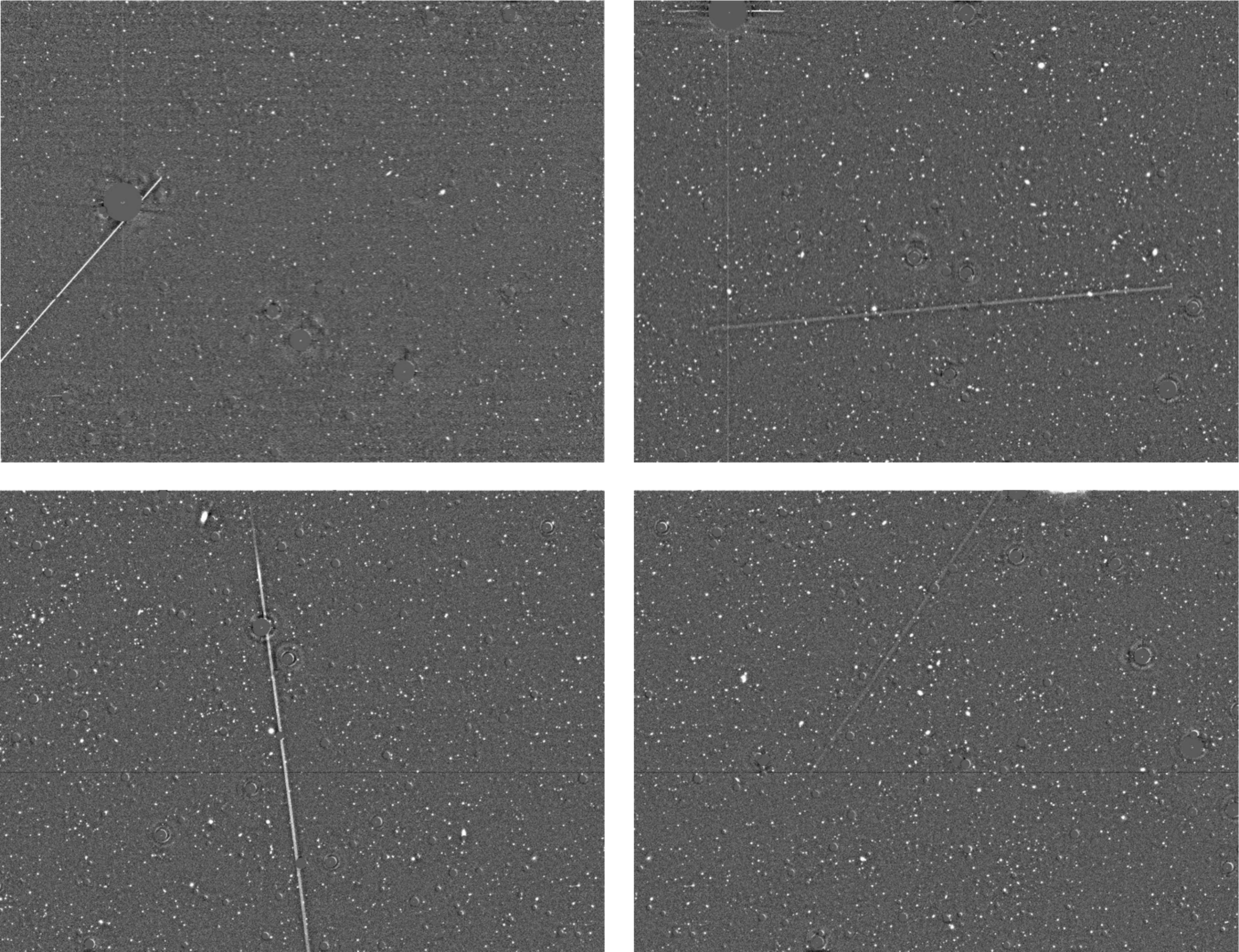}
\bigskip
\begin{minipage}{12cm}
\caption{Montage showing streaks produced by four uncorrelated objects detected by the ILMT. The individual images cover 29 x 22 arcmin and have had bright stars removed. \label{fig:uncorrelated}
}
\end{minipage}
\end{figure}

The correlated objects have estimated magnitudes as faint as 14.5 and were detected with a signal-to-noise ratio greater than 200. The flux of the faintest uncorrelated object is 6.4 times smaller than that of the faintest correlated object. Four examples of uncorrelated objects are shown in Fig.\,\ref{fig:uncorrelated}.

These results indicate an average rate of 9.2 detectable objects per square degree per hour passing near the zenith at the latitude of the ILMT. Objects having orbital inclination less than the latitude never reach the zenith and are thus not counted. One therefore expects that the rate will be higher at lower latitudes.

\begin{table}
\centering
\begin{minipage}{88mm}
\caption{Observations and results. \label{tab:results}
}
\end{minipage}
\bigskip

\begin{tabular}{ll}
\hline
\textbf{Parameter} & \textbf{Value}  \\
\hline
Area of sky observed, per image &		0.139 sq. deg. \\
Number of images examined &		515 \\
Total observing time &			65.1 hr \\
Number of streaks detected &	83 \\
Number correlated with catalogued objects &	43 \\
Number of uncorrelated objects & 		40 \\
Percent uncorrelated &		48\% \\
Magnitude range of correlated objects &		6.9 - 14.5 \\
Altitude range of correlated objects &		451 - 25,014 km \\
Satellite fraction &			63\% \\
Rocked body fraction &		17\% \\
Debris fraction &			20\% \\
\hline
\end{tabular}
\end{table}

\section{Discussion}

These initial results demonstrate the potential of the ILMT for the serendipitous detection of satellites and orbital debris. Almost half of the objects that we detected could not be matched to catalogued objects, suggesting a significant level of incompleteness, or inaccuracy, of TLEs in the public database.

Streaks from bright objects are a concerning source of contamination for astronomical observations. Many of the streaks produced by the brighter objects that we detected are as wide as 12 arcsec and essentially obliterate faint stars and galaxies that they cross. Subtraction of the streaks is problematic due to the high levels of photon noise that they produce, and intrinsic variations in brightness as the objects rotate \citep{Tyson_2020}.

These results were obtained from just 10 nights of ILMT observations, taken during commissioning, where roughly 40\% of the time was used for engineering tests and adjustments and therefore not available for observations. In regular operation, the ILMT is expected to obtain on the order of 1500 clear dark hours of observations per year, which is roughly 20 times the present data set. This will provide a unique opportunity to monitor the increasingly-crowded space environment.

\begin{acknowledgments}
The 4m International Liquid Mirror Telescope (ILMT) project results from a collaboration between the Institute of Astrophysics and Geophysics (University of Li\`{e}ge, Belgium), the Universities of British Columbia, Laval, Montreal, Toronto, Victoria and York University, and the Aryabhatta Research Institute of observational sciencES (ARIES, India). The authors thank Hitesh Kumar, Himanshu Rawat, Khushal Singh and other observing staff for their assistance at the 4m ILMT.  The team acknowledges the contributions of ARIES's past and present scientific, engineering and administrative members in the realization of the ILMT project. JS wishes to thank Service Public Wallonie, F.R.S.--FNRS (Belgium) and the University of Li\`{e}ge, Belgium for funding the construction of the ILMT. PH acknowledges financial support from the Natural Sciences and Engineering Research Council of Canada, RGPIN-2019-04369. PH and JS thank ARIES for hospitality during their visits to Devasthal. BA acknowledges the Council of Scientific $\&$ Industrial Research (CSIR) fellowship award (09/948(0005)/2020-EMR-I) for this work. MD acknowledges Innovation in Science Pursuit for Inspired Research (INSPIRE) fellowship award (DST/INSPIRE Fellowship/2020/IF200251) for this work. TA thanks the Ministry of Higher Education, Science and Innovations of Uzbekistan (grant FZ-20200929344). This work is supported by the Belgo-Indian Network for Astronomy and astrophysics (BINA), approved by the International Division, Department of Science and Technology (DST, Govt. of India; DST/INT/BELG/P-09/2017) and the Belgian Federal Science Policy Office (BELSPO, Govt. of Belgium; BL/33/IN12). We thank Bikram Pradhan for helpful comments on the manuscript. 
\end{acknowledgments}

\begin{furtherinformation}

%\begin{orcids}
%\orcid{0000-1111-2222-3333}{Hàrry}{Harrisòn}
%\orcid{1111-2222-3333-4444}{Leonie}{van Leon}
%\orcid{2222-3333-4444-5555}{Lotta}{Lothardis}

%{\sl This section is optional.
%You may list here the ORCIDs of those authors who would like to share them, one per line, with the \verb|\orcid{|\texttt{\emph{ORCID}}\verb|}{|\texttt{\emph{First name}}\verb|}{|\texttt{\emph{Last name}}\verb|}| command.
%This command typesets the information, and makes the ORCIDs themselves active links to the corresponding records on \href{https://orcid.org}{orcid.org}.

%Unlike in this sample, no other text should actually be included here and this section should reduce to a bare list.
%The \verb|\orcid| command controls line feeds by itself; please do not insert any \verb|\\| or \verb|\newline| before or after them.}
%\end{orcids}

\begin{authorcontributions}
%This section is mandatory when there is more than one author.
%The contributions of each author (identified by their initials) must be declared.
%We recommend to follow the \href{http://credit.niso.org}{CRediT} taxonomy (Contributor Roles Taxonomy).
This work results from a long-term collaboration to which all authors have made significant contributions.
\end{authorcontributions}

\begin{conflictsofinterest}
%This section is \emph{mandatory}.
%Authors must declare any personal or professional circumstances that may be perceived as influencing the research reported in the paper.
%If there is no conflict of interest, please state that 
The authors declare no conflict of interest.
\end{conflictsofinterest}

\end{furtherinformation}

\bibliographystyle{bullsrsl-en}

\bibliography{S11-P13_HicksonP}

\end{document}